\documentclass[12pt]{article}
\usepackage{amsmath}
\usepackage{graphicx,psfrag,epsf}
\usepackage{enumerate}
\usepackage{natbib}
\usepackage{url} 

\usepackage{adjustbox}
\usepackage{array}

\usepackage{booktabs}

\newcolumntype{R}[2]{%
    >{\adjustbox{angle=#1,lap=\width-(#2)}\bgroup}%
    l%
    <{\egroup}%
}

\usepackage{hyperref}
\hypersetup{
    citecolor=black,
    colorlinks=true,
    linkcolor=blue,
    filecolor=magenta,      
    urlcolor=black,
}

\newcommand{\blind}{0}

\newcommand{\R}{{R} }

\begin{document}

\def\spacingset#1{\renewcommand{\baselinestretch}%
{#1}\small\normalsize} \spacingset{1}



\if0\blind
{
  \title{\bf A Review of Containerization for Interactive and Reproducible Analysis}
  \author{Gregory J. Hunt\hspace{.2cm}\\
    Department of Mathematics, William \& Mary\\
    and \\
    Johann A. Gagnon-Bartsch \\
    Department of Statistics, University of Michigan}
  \maketitle
} \fi

\if1\blind
{
  \bigskip
  \bigskip
  \bigskip
  \begin{center}
    {\LARGE\bf }
\end{center}
  \medskip
} \fi

\bigskip
\begin{abstract}

In recent decades the analysis of data has become increasingly computational. Correspondingly, this has changed how scientific and statistical work is shared. For example, it is now commonplace for underlying analysis code and data to be proffered alongside journal publications and conference talks. Unfortunately, sharing code faces several challenges. First, it is often difficult to take code from one computer and run it on another. Code configuration, version, and dependency issues often make this challenging. Secondly, even if the code runs, it is often hard to understand or interact with the analysis. This makes it difficult to assess the code and its findings, for example, in a peer review process. In this review we describe the combination of two computing technologies that help make analyses shareable, interactive, and completely reproducible. These technologies are (1) analysis containerization, which leverages virtualization to fully encapsulate analysis, data, code and dependencies into an interactive and shareable format, and (2) code notebooks, a literate programming format for interacting with analyses. The fusion of these two technologies offers significant advantages over using either individually. This review surveys how the combination enhances the accessibility and reproducibility of code, analyses, and ideas.

\end{abstract}

\noindent%
\vfill

\newpage
\spacingset{1.45} 

\section{Introduction}
\label{sec:intro}

Before the widespread adoption of peer-reviewed scientific journals, it was not uncommon for scientists to keep their findings secret. Famously, Leonardo Da Vinci wrote in mirrored handwriting to obfuscate his notebooks and Isaac Newton kept hidden his development of calculus for nearly forty years \citep{NationalAcadem2009}. Modern science, however, advances through a rich process of open and timely sharing. Today, there are a plethora of ways to share results such as talks at conferences, proceedings, seminars, posters, peer-reviewed literature and pre-print repositories. Open sharing not only allows results to be disseminated and built upon, but also allows scrutiny and verification of the research and is fundamental to the scientific process itself. However, as scientific analysis has progressed, so too has the notion of sharing. In particular, the last several decades have seen the analysis of scientific data become heavily computational. This is especially true of statistical work, where coding has become deeply intertwined with statistical analysis. Correspondingly, the notion of what it means to share research results has also expanded \citep{Ellis2018}. The modern notion of sharing research encompasses not only sharing prose and proofs, but also sharing code and data. 


It is now commonplace for data and the accompanying analysis code to be shared through online repositories. Indeed, many peer-reviewed journals require it. For general purpose code, a popular sharing platform is github \citep{github}. Language specific repositories for software packages also exist, e.g. CRAN for R packages \citep{cran} or PyPI for Python \citep{pypi}. Moderately sized datasets may also be hosted on github or Kaggle \citep{kaggle}. This open sharing of analysis code is a growing trend in statistics. Nonetheless, it faces several practical challenges. Among these, two important issues are (1) actually running the shared code, and (2) understanding and interacting with the code. 

The first challenge is that code that runs on one computer may not always run on another. For example, the package may not be available for the current version of the language or dependencies of the package may fail to install. Modern analysis often relies on a large and complex collection of interdependent software packages and thus there are many places for such version or dependency issues to arise. Similarly, directory structures across machines may not be identical and, for example, data, code, or other files may not reside where the analysis is expecting. Fixing such problems often entails a significant investment of time and energy. For example, troubleshooting failed installations of dependencies can often lead down a chain of fixing cryptic installation errors which is difficult even for an experienced user. 

In addition to the challenges of taking analysis from one computer and running it on another, a second major challenge is difficulty understanding or interacting with code. While it may be impractical or unnecessary to insist on understanding code on a line-by-line basis, a lot can be learned about an analysis by making small modifications to code. For example, one can explore different parameter settings or function arguments and see how output changes. Here, simply sharing raw code is often inadequate. Moderately complicated code, even well-written code, can be difficult to understand and explore. Consequently, it is often difficult for third-parties to find reasonable entry-points into code to modify or scrutinize the analysis.  

These issues limit gaining a deeper understanding of shared of statistical and scientific results. In this work we will review how two computational toolsets can be combined to help address these problems. These toolsets are: (1) analysis containerization, and (2) interactive notebooks. Containerization is a virtualization technology that allows encapsulation of an entire computing environment including data, code, dependencies, and programs into a reproducible, shareable, and self-contained format \citep{Nust2020}. When a third party takes the container and runs it on their own computer, it will be as if they are instead working in the computational environment where the analysis was originally done. All of the programs, files, code, data, and configurations will be exactly reproduced as they were in that original environment. While any format or organization of analysis can be containerized, this approach has proven to be particularly user-friendly when used to virtualize interactive code notebooks. Code notebooks are an increasingly popular approach to analysis that allow natural interweaving of commentary, code, and output. Containerizing code notebooks makes for some of the most clear, concise, and intuitive ways of documenting and interacting with code and for providing a user-friendly interface to analysis.

In this work we will review how the merging of containerization and notebook software can be used to create interactive and reproducible analyses. In addition to an overview, we will also make concrete recommendations for what we believe to be the most straight-forward tools and workflows for enhancing reproducibility through containerized notebooks. The remainder of this paper is organized as follows. Section~\ref{sec:container} reviews barriers to computational reproducibility in statistics, how containerization helps, and the landscape of available tools. Section~\ref{sec:notebook} reviews interactive notebooks and surveys a selection of software options for writing notebooks that can be  easily containerized. Finally, section~\ref{sec:concl} concludes with a discussion of additional benefits of containerization and notebooks.

\section{Containerizing Analyses}
\label{sec:container}

\begin{figure}
    \centering
    \includegraphics[width=\textwidth]{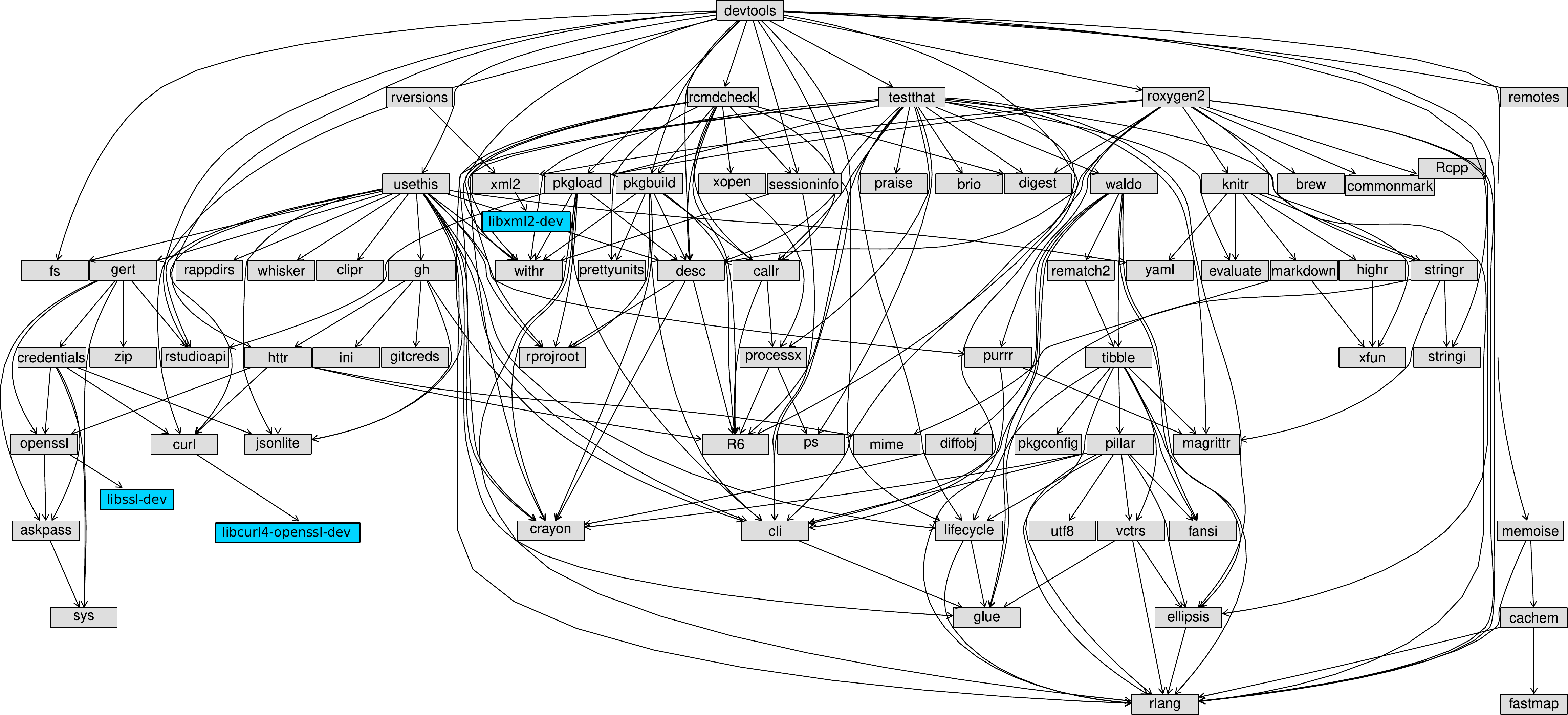}
    \caption{Dependency graph for the \R package {\tt devtools}. Grey boxes are \R add-on packages. Arrows indicate dependency. Blue boxes are system-level dependencies of packages for Linux OS Ubuntu ver. 20.04.}
    \label{fig:dep}
\end{figure}

The basic computational reproducibility problem is that often code encounters errors when moved from one computer to another. This was emphasized by the American Statistical Association's 2017 recommendations on reproducible research, which noted that ``[reproducible code] may initially sound like a trivial task but experience has shown that it’s not always easy to achieve this seemingly minimal standard.''\citep{asarepro} One major source of trouble is ensuring correct code dependencies. The most familiar example of this is installing add-on packages for a language like {\tt ggplot2} for R or {\tt numpy} for python. While add-on dependencies are easy to install in some cases, this can quickly become complicated, for example, if the original analysis used a now out-of-date version of a package. Furthermore, add-on packages often have their own dependencies. Thus installing a single package may actually require a large network of interrelated packages to be configured. Figure~\ref{fig:dep} displays the package dependency network for the popular \R package {\tt devtools} which has 72 add-on package dependencies and 3 system-level library dependencies \citep{devtools}. 

There has been significant effort in the R community to address some of these add-on dependency issues. The CRAN task view on reproducible research lists several packages for this purpose like {\tt checkpoint}, {\tt groundhog}, and {\tt renv} \citep{cranrepro,checkpoint,groundhog,renv}. These tools all enhance reproducibility by maintaining a local archive of packages as used at the time of analysis. This archive can subsequently be distributed with analysis code so that the correct add-on versions are available to third-parties. While \R has such archival packages available, other languages have comparatively less support. Furthermore, analyses often have dependencies beyond simple add-on packages that cannot be archived in this way. For example, code typically depends on programming language and operating system versions, and system-level library code (as in Figure~\ref{fig:dep}).

Recently, some have sought to solve these broader dependency issues using virtualization, a well-studied software engineering solution to dependency problems \citep{Nust2020, Olaya2020}. Virtualization encapsulates code and all of its dependencies into a virtual computing environment that can be easily disseminated. One can think of virtualization as making a copy of the computer where the code was originally written. This virtual copy can be taken to another computer and run with little to no setup or configuration. While virtualization has been around for decades, containerization is the latest incarnation of the technology and comes with several key advantages over its predecessors. Previous technology virtualized the entire computer from hardware on up. This meant that virtualization was resource intensive and slow to use. Conversely, containerization is incredibly light-weight. Containers only virtualize the high-level components of the operating system (e.g. code, configuration files, software and data) and seamlessly re-use the stable low-level processing components of the host operating system \citep{dockerbook}. Indeed, starting up a container doesn't actually start up a second instance of an operating system; it largely just changes all references for resources, system libraries, files, and data, to refer to a particular isolated section of the computer. The light-weight nature of such containers means that the resource footprint is small making them quick to upload, download, and share. Furthermore, since starting a container largely just changes the references to resources in the environment, containers are user-friendly, start up nearly instantaneously, and run code at speeds nearly identical to the host computer \citep{Felter2015}.

\subsection{Containerization in Practice}\label{sec:run}

\begin{figure}
    \centering
    \includegraphics[width=\textwidth]{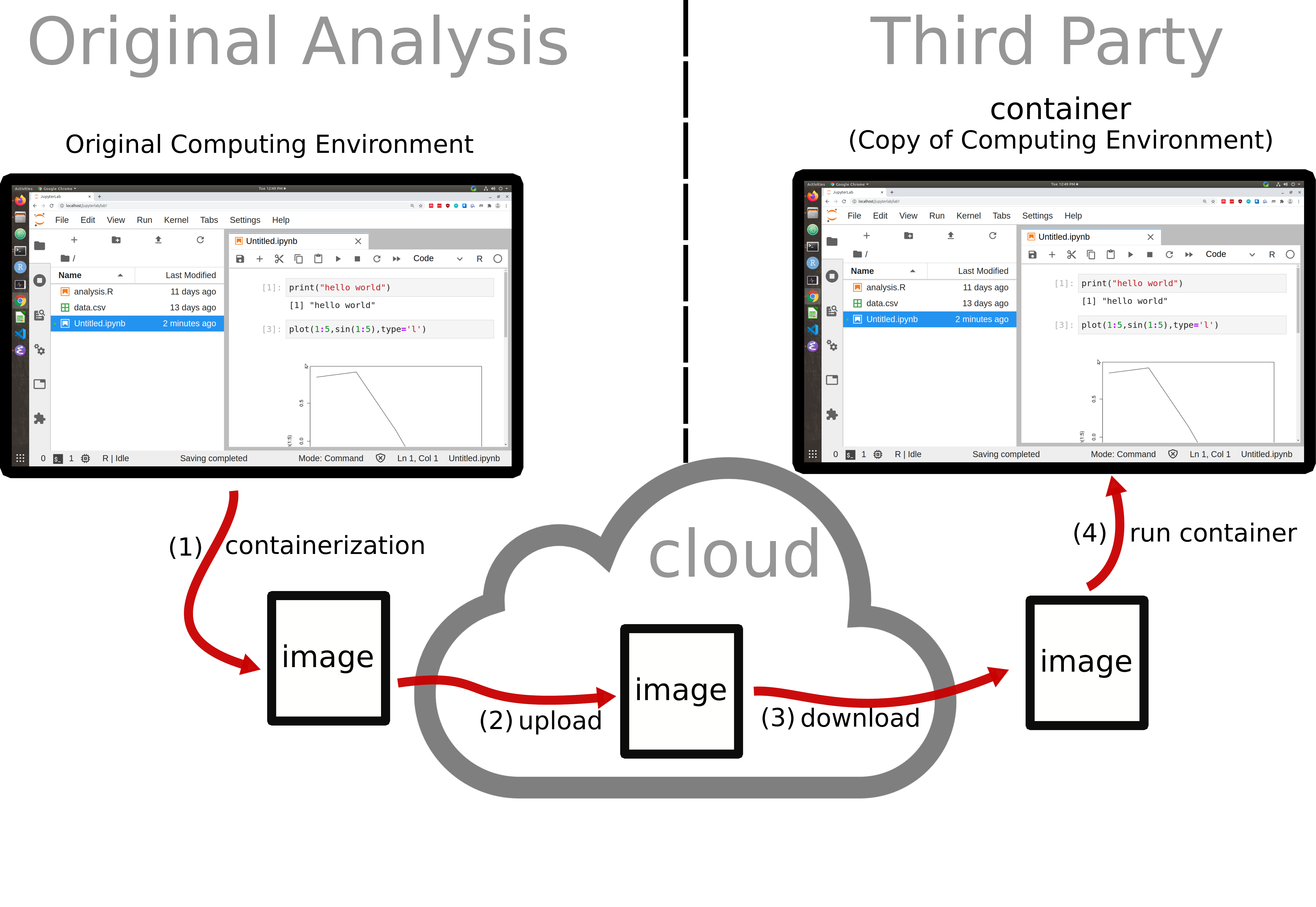}
    \caption{Typical sharing of containerized analysis. (1) The computing environment is containerized, creating a self-contained image file. (2) This image file may be uploaded to the cloud and then (3) downloaded by a third party. (4) From there, the third party may use the image to re-create the original computing environment. }
    \label{fig:workflow}
\end{figure}

\begin{figure}\centering
  \includegraphics[width=\textwidth]{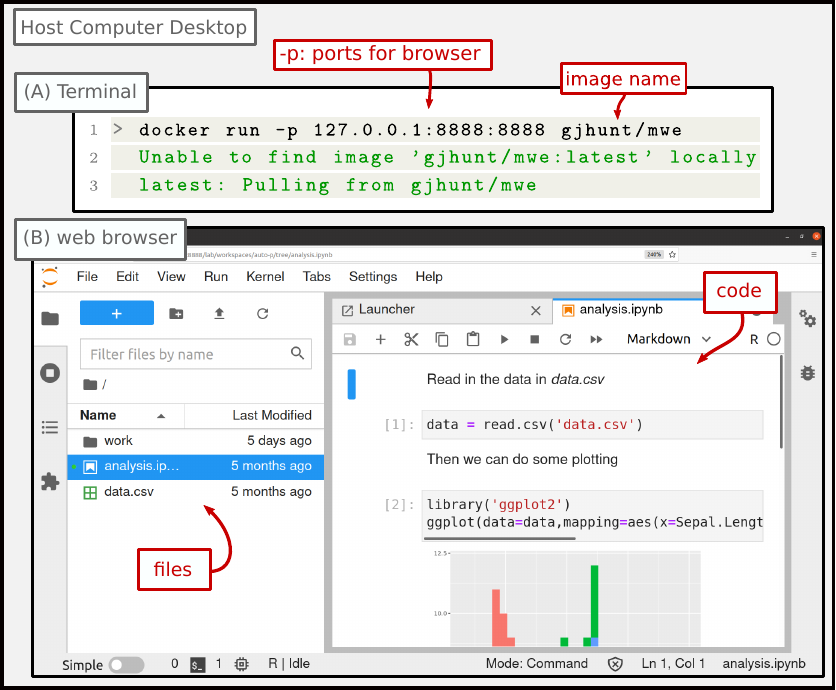}
\caption{Example of interacting with a containerized analysis. (A) Here we use the containerization software Docker to launch the container. The command {\tt docker run} starts the container. The flag {\tt -p} allows us to specify the port forwarding to we can interact using the web browser. (B) The container may now be interacted with through the web browser on the host computer via a graphical interface running from the container. Here, the interface is the Jupyter lab IDE. The container has the necessary data and code files and an installation of \R to run the analysis through this web interface. While the end-user may interact naturally with the analysis through a web-browser on the host computer, all of the code, files, and software reside in the container's pre-configured environment.} 
\label{lst:pull}
\end{figure}

\begin{figure}
  \includegraphics[width=.8\textwidth]{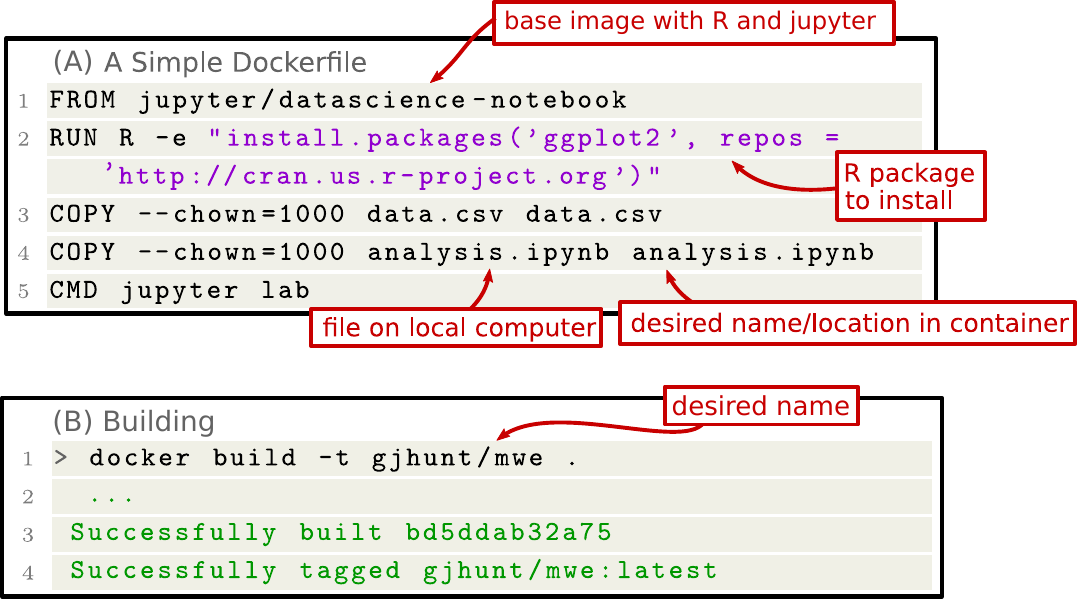}\centering
  \caption{(A) Example configuration file for building an image using Docker. {\bf (Line 1)} {\tt FROM} specifies the base image named {\tt jupyter/datascience-notebook} to get a container with R and jupyter. {\bf (Line 2)} {\tt RUN} executes code which calls R and installs {\tt ggplot2}. {\bf (Line 3-4)} Copies the data and code. First argument to {\tt COPY} is location on host, second argument is desired location in container, the flag {\tt --chown} sets the ownership of the file to the container's user. {\bf (Line 6)} {\tt CMD} sets the command executed when the container starts, here starting jupyter lab. (B) Building image from Dockerfile. Flag {\tt -t} specifies the image name as {\tt gjhunt/mwe}. ``.'' specifies necessary files to copy are in the current directory. }
  \label{lst:build}
\end{figure}

Containerization has been an increasingly adopted tool for reproducibility widely across the scientific community including areas such as geography, psychology, environmental science, metagenomics and many others \citep{Knoth2017, Wiebels2021, Essawy2020, Visconti2018, Nust2020, Olaya2020}.
To set the stage for a review of containerization technology we will first illustrate how containerization is used in practice. We will present an archetypal example of containerizing and sharing an analysis from three different perspectives: (1) the high-level view of sharing containerized analyses, (2) the end-user experience of interacting with a third-party containerized analysis, and (3) the first-party task of containerizing an analysis for dissemination. These will correspond to Figures~\ref{fig:workflow}, \ref{lst:pull}, and \ref{lst:build}, respectively.

Figure~\ref{fig:workflow} displays a high-level overview of how statistical analyses are containerized and shared. 
First, the entire computing environment in which the analysis was originally run is encapsulated into a single file. This file, called an image, is essentially a copy of the system on which the analysis was conducted. The image file may be shared, for example, by uploading it to the cloud. From there, the image may be downloaded by a third party and, with just a few keystrokes, the third party is placed into a duplicate of the original computing environment (called a ``container''\footnote{An ``image'' refers to the actual file that may be uploaded, downloaded or shared, while a ``container'' refers to an ephemeral instance running on the computer.}). All of the data, code, dependencies, configurations and software are precisely set up as in the original environment, and thus set up to reproduce the analysis exactly. The goal of containerization is to ensure that if the code worked when containerized, it will work when the image is run by a third party. This figure emphasizes that containerizing and sharing analyses is a simple process akin to uploading code to github. However, unlike uploading code to github, containerizing analyses ensures exact computational reproducibility and enables natural interaction with the shared analyses.

Figure~\ref{lst:pull} shows in detail what using a shared containerized analysis looks like from the viewpoint of an end-user. First, the container is downloaded and started with a single command. The default interface to a container is through the command line. However, as shown in Figure~\ref{lst:pull}, when combined with code notebook software the analysis is accessible via an interactive graphical interface through the computer's web browser. Alone, the containerization ensures exact reproducibility but is not user-friendly. Conversely, notebook software alone provides a user-friendly environment but doesn't guarantee exact reproducibility. The combination achieved by containerizing notebooks gets the best of both worlds. We will more fully explore this in Section~\ref{sec:notebook}.

We can see from Figure~\ref{lst:pull} that the container's environment contains all of our files necessary for analysis including data and code scripts. However, in addition to merely allowing inspection of the data or scripts, the container also comes with an installation of \R so that the user can actually run the code and analysis through the interactive notebook interface. It is important to keep in mind that while the end-user accesses the container and its contents through the web browser on the host computer, the data, code, software installations and back-end to the interface all actually reside in the container. The web browser merely provides a window into the running container through which one may use the tools installed in the container and interact with the code and data it contains. Indeed, none of these programs or files need be installed on the host computer in order to use the web browser to interactively access the versions running in the container. This is the power of sharing containerizing analyses -- it allows users to bring to bear the full power and convenience of popular graphical interfaces to fully encapsulated analysis environments with a single command. 

To set up an image a configuration file must be written giving instructions of which files and programs to be copied and installed. Images need not be built from the ground up but, instead, one can simply add on to existing pre-configured images to create new ones. In section~\ref{sec:cccontainer} we will discuss the panoply of pre-configured ``base'' images available in online image repositories. Such repositories make containerizing analyses simple as one can choose a nearly-complete image, with desired software like Jupyter lab and \R already installed, and simply add a small amount of project-specific code, data, and documentation. 

Figure~\ref{lst:build}~(A) displays the configuration file used to build the image from Figure~\ref{lst:pull}. In five lines the configuration specifies a base image with \R and Jupyter already installed, installs a desired \R add-on package, copies over data and analysis code, and starts the Jupyter lab interface. Such a simple configuration file is quite typical for containerizing statistical analyses. Most of the heavy lifting is done by the base image which sets up a nearly complete environment. On top of this base image one needs only to install the necessary software packages or language add-ons and copy over the data and code.
Once the configuration file has been written, the image needs to be built, after which, it may be run or shared.  Building the image is illustrated in Figure~\ref{lst:build}~(B). 

\subsection{The Containerization Landscape}\label{sec:cccontainer}

While virtualization can trace its roots all the way back to early mainframe computers, modern lightweight containerization was largely popularized with the software Docker starting in 2013 \citep{Graziano2011,docker}. While other tools have been developed since then, the present space of user-friendly containerization software for statisticians and scientists has two major players: (1) Docker, and (2) Singularity \citep{docker,singularity}. In the remainder of this section we will briefly compare these options, summarizing findings in Table~\ref{tbl:cc}. 

Portability is paramount to reproducibility. Docker and Singularity are both free and open source and built off a Linux base. Consequently, they both work on Linux. However Singularity does not have native support on Windows or MacOS while Docker has both support and a graphical interface for these systems. Nonetheless, Singularity is partially inter-operable with Docker and can run Docker images or use them as a base image. Conversely, Docker can only work with Docker images.

A significant distinction is that Docker requires administrator privileges to run, while Singularity does not. This makes Singularity capable of deploying software on high-performance computing clusters where users do not have these rights. If one wishes to run Docker on a cluster they may consider using Podman instead. Podman is a re-implementation of Docker that doesn't require administrative privileges \citep{podman}. Podman is only available on Linux.

In addition to required privileges, there are differences in system isolation. Singularity does not by default isolate the host computer's file-system or network interface from the container while Docker does. This makes Singularity's default behavior less secure for running unverified third-party analyses but more amenable for deploying non-interactive code to clusters. Singularity's default configuration also locks containerized analyses as read-only unlike Docker. This makes it relatively difficult explore and edit third-party analysis code with Singularity.

A significant feature of containerization is the vast wealth of base images off of which to base a containerized analysis. The most popular repository is the official Docker repository Dockerhub with more than 100,000 images freely available and usable by all three of Docker, Singularity and Podman \citep{dockerhub}. Users may freely upload/download images to/from Dockerhub which is useful for sharing but not the ideal for long-term archival of data-heavy images. As an alternative to uploading images to Dockerhub, we suggest sharing links to redundant copies of images on cloud-based storage. Other public image repositories also exist like Biocontainers for analyzing biological data \citep{bioc}.

Table~\ref{tbl:cc} summarizes this discussion. For containerizing shareable and reproducible analyses we recommend Docker as it is the most widely used containerization software with cross-platform support, a user-friendly interface, and a huge ecosystem of base images off of which one may build. Nonetheless, for deploying containerized analyses to high-performance computing environments Singularity and Podman have substantial strengths. 

\begin{table}\centering
  \begin{tabular}{@{}llll@{}}
\toprule
&{\bf Docker}& {\bf Singularity}& {\bf Podman}\\
\midrule
O/S Support& Linux, Mac, Windows& Linux& Linux\\
Image Type Support& Docker& Docker, Singularity& Docker\\
Admin. Privileges & Required& Not Required& Not Required\\ 
Host/Container Isolation& Yes& No& Yes\\
Container Mutability& Read/Write& Read Only& Read/Write \\
\bottomrule
\end{tabular}
  \caption{Comparison of Docker, Singularity, and Podman for containerization of reproducible analyses.}
  \label{tbl:cc}
\end{table}

\section{Notebooks and Interactivity} \label{sec:notebook}

\begin{figure}
    \centering
    \includegraphics[width=\textwidth]{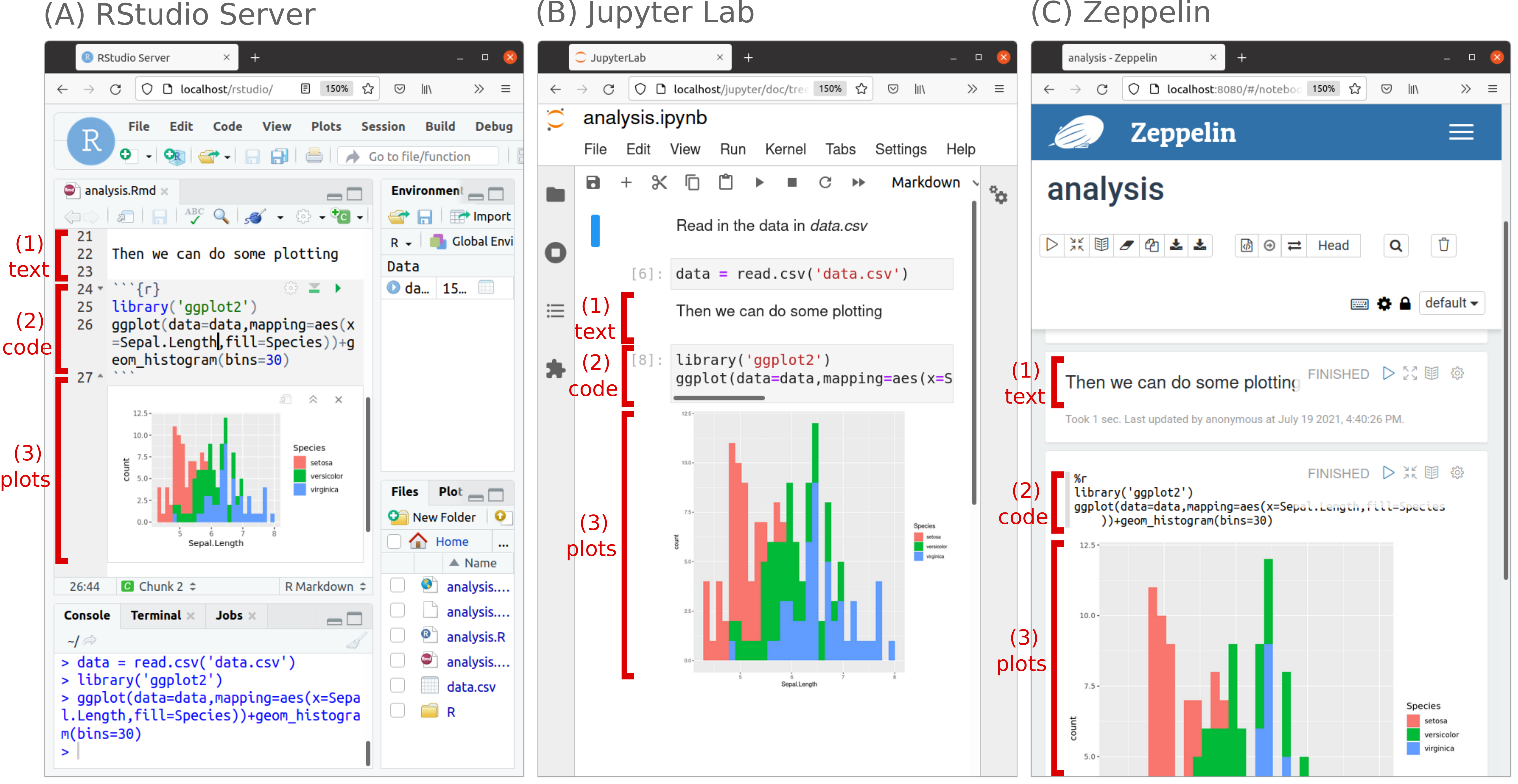}
    \caption{Interactive notebook environments run through the web browser using (A) RStudio Server, (B) Jupyter Lab, and (C) Zeppelin. While different notebook formats and software tools exist, all notebooks share the feature of organizing analysis as a sequence of chunks of (1) text or (2) code and its associated (3) output. }
    \label{fig:notebooks}
\end{figure}

Notebooks are a document format that allow interweaving of commentary, code, and output all together. Figure~\ref{fig:notebooks} displays three examples of popular code notebook formats which will be reviewed in section~\ref{sec:ccnotebook}. While there are several variants of code notebooks, they all structure analysis as a sequence of ``chunks'' that can be edited and evaluated one at a time. Each chunk can either be text or code. Text chunks typically allow markdown formatting while running code chunks displays the output inline.

This chunked notebook structure naturally breaks up code and annotation into bite-size segments which helps organize analysis into a logical flow. The ability to interweave bite-sized chunks of code and text makes notebooks well suited for explaining, showcasing, and exploring data and analysis. Since chunks can be edited and run one at a time, each chunk provides a natural entry-point into a small portion of the analysis. For example, one can pick a segment of the analysis they wish to explore, edit the chunk of code, run it, and observe the subsequent change in output. This allows one to experiment with small changes to code, e.g., testing different tuning parameters or optional arguments to functions, and immediately observe the changes to local output without having to re-run the entire analysis. This provides a natural way to play with code in order to build up an understanding of how the code works and test the robustness of the analysis to alterations.

Notebook software can also make containerizing a fully interactive analysis user-friendly. By default, a containerized analysis requires the user to interact with the code entirely through the command line. However, if we containerize notebook software in addition to the code, data, and other dependencies, then we can bring the full power of popular coding environments to our containerized analysis. This is particularly easy if the notebook software is accessible through a web browser. In this case, we can run the notebook back-end from within the container but access the interactive computing interface from the host computer's browser. This is illustrated in both Figure~\ref{lst:pull} and Figure~\ref{fig:notebooks}. This combination is the best of both worlds as it brings the native feel of doing analysis on one's own computer to completely self-contained and reproducible analyses. 

\subsection{Options for Interactive Notebooks}\label{sec:ccnotebook}

There are three major notebook types that are simple to containerize: (1) RStudio, (2) Jupyter, and (3) Zeppelin. Aside from these three, there is other proprietary notebook software like Wolfram Mathematica or MATLAB Live Scripts; however these are closed-source and difficult to containerize. Conversely, third-party software like Pycharm or VSCode can write and run Jupyter notebooks but are more complicated to containerize as they lack a native web interface. Consequently, this section compares Rstudio, Jupyter, and Zeppelin, all three of which have an easily containerizable web interface along with official images and support on Dockerhub. A summary of this comparison is presented in Table~\ref{tbl:ccnotebook} and examples of the software interfaces are illustrated in Figure~\ref{fig:notebooks}.

All three notebook options support a large array of languages popular for data science like R, Python, Julia, Octave, and many others. RStudio boasts over 55 language interpreters, Jupyter lists over 150, and Zeppelin has support for 37 (with a focus on languages for clusters like Hive, Pig, Spark and BigQuery). Zeppelin can also create chunks that effectively run RStudio or Jupyter as a backend and thus directly borrow the features and languages they support. While Jupyter requires that all code chunks in a notebook use the same language both RStudio and Zeppelin allow notebooks to mix and match languages across chunks. Furthermore, RStudio has extensive support for {\tt reticulate}, which allows analysis using both R and Python at the same time in a shared computing environment. 

An important distinction is the format of the notebook file and how it interacts with third-party software.  Jupyter and Zeppelin serialize the notebook and save it in a single densely encoded file. Conversely RStudio saves input code and markdown in a plain-text ``R Markdown'' file and renders output into a separate HTML file. Each format has its advantages and drawbacks. The encoding used by Jupyter/Zeppelin allows them to save output text or plots in one file alongside input code, commentary, and rendered markdown. This is useful for showcasing results because, unlike a traditional code scripts, the notebook has output embedded and one need not re-run the code to view the results. However the file format is densely encoded which can cause difficulties when combined with other software. In particular, one cannot easily track changes to these notebooks in a human-readble format using version control software like git since small changes to output can prompt a cascading change to hundreds of lines of the dense encoding.

Alternatively, RStudio uses a combination of plain-text R Markdown for input and HTML for rendered output. An advantage of such an approach is that the input code and commentary are saved in a human-readable format which is more versatile for editing by general software and can be meaningfully tracked by version control schemes. The shareable HTML rendering also contains an embedded downloadable copy of the R Markdown file if one wishes to download the underpinning R Markdown code. In addition to rendering to HTML, RStudio allows rendering notebooks into a host of output types for display like PDF, Word or Powerpoint. Jupyter can also render its notebooks into to a slightly smaller selection of similar display formats however Zeppelin does not have such support. Despite these format differences, from the viewpoint of interacting and exploring analyses all three of RStudio, Jupyter and Zeppelin have broadly similar behavior and allow users to edit and run code chunks one at time, viewing output in-line in the editor. 

To allow conversion between formats Jupyter has the Jupytext plugin which allows one to conduct analysis using Jupyter but maintain a simultaneous synchronized version in R markdown or as a simple executable script. This allows the best of both Jupyter and R Markdown notebooks, and in
particular makes Jupyter compatible with version control software. Zeppelin only supports converting their notebooks into the Jupyter format while RStudio does not natively support conversion of R Markdown to other formats, nonetheless Jupytext can enable this conversion.

All three of Jupyter, RStudio and Zeppelin have the ability to embed interactive widgets into notebooks using popular interactive libraries in languages like R and Python. As embedding widgets typically takes extra configuration it makes a strong case for containerization which will ensure the back-end software is correctly set up to support such interactivity. While Jupyter has support for interactive elements in both Python and R, RStudio primarily supports these through its R Shiny platform for building web apps. As Zeppelin can create notebook chunks running the backend language interpreters of both RStudio (including R Shiny) and Jupyter it can create notebooks that naturally embed interactive R Shiny apps or Jupyter widgets. Zeppelin also has its own interactive visualizations backend via Apache Spark.

A common challenge when using notebooks is that chunks need to be run sequentially and so to explore chunks later in the analysis one needs to run earlier time-intensive code. To facilitate entering the analysis at arbitrary points it is good practice to save the output of time-intensive chunks. This allows subsequent chunks to simply load the pre-computed intermediate results instead of requiring a preceeding time-intensive chunk be run. This is practice, called results caching, can be done manually by reading/writing serialized objects to/from disk e.g. using {\tt pickle} in Python or {\tt read/writeRDS} in R. Containers are well-suited for this as one can distribute notebooks together with cached results.

\begin{table}\centering
  \begin{tabular}{@{}llll@{}}
    \toprule
    & {\bf RStudio}& {\bf Jupyter}& {\bf Zeppelin}
    \\ \midrule
    Notebook Type& R Markdown& Jupyter& Zeppelin\\
    & (text, HTML)& (JSON)& (JSON)\\
    Convertible to & None& R Markdown, code script& Jupyter\\
    Language Support& $\geq$56& $\geq$152& $\geq$37\\
    Export Types& $\geq$18& $\geq$9& None\\
    Widgets Backends& Shiny,& Several& Jupyter,\\
    &HTML Widgets& &Shiny \\
    Caching Support& R (native)& Python (add-ons)& None\\
    Ex. Docker Image& {\tt rocker/rstudio}& {\tt jupyter/base-notebook}& {\tt apache/zeppelin} \\
\bottomrule
  \end{tabular}
  \caption{Comparison of notebook software options RStudio, Jupyter, and Zeppelin.}
  \label{tbl:ccnotebook}
\end{table}

\begin{figure}
    \centering
    \includegraphics[width=\textwidth]{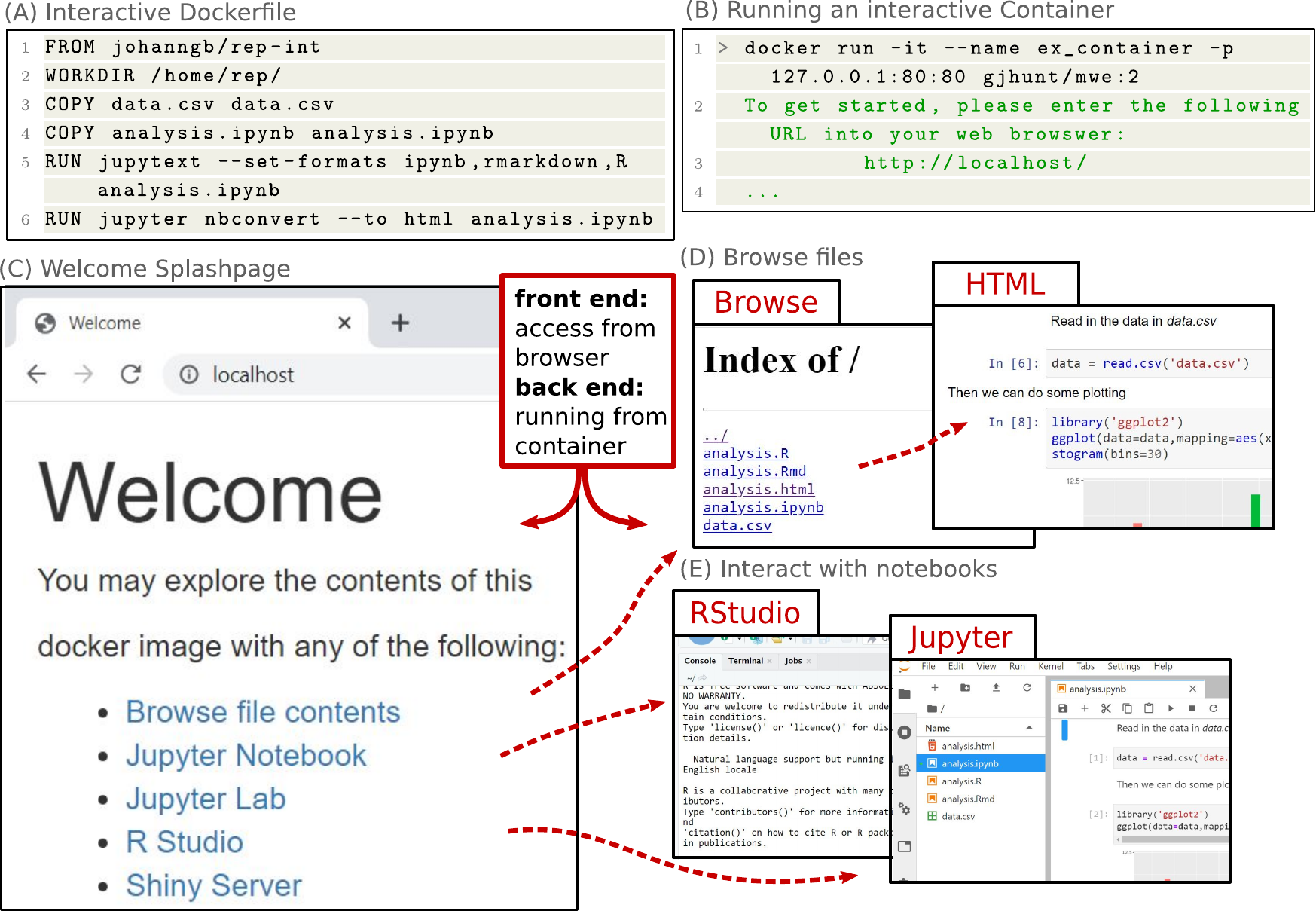}
    \caption{An interactive containerization workflow. (A) {\bf (Line 1)} An interactive dockerfile built from {\tt johanngb/rep-int} base. {\bf (Line 5)} {\tt jupytext} links the Jupyter notebook to a R Markdown notebook and script.  {\bf (Line 6) } {\tt jupyter} runs the notebook and saves input/output as a HTML document for showcasing. (B) We build the image and name it adding the tag {\tt :2} to indicate it is version 2 of our previous example. Subsequently, we may run the image interactively with {\tt -it}, naming it with {\tt --name} and correctly mapping ports with {\tt -p}. (C) The start page for the interactive container. Several options for interacting with the analysis files are listed. (D) We may browse the files or (E) open the notebooks with one of several choices of graphical web-based interfaces running from the container. }
    \label{fig:dockerint}
\end{figure}

In summary, for every-day statistical analyses we recommend either Jupyter or RStudio but also using Jupytext to mirror copies into both formats. Nonetheless, if one needs to connect to cluster architecture Zeppelin likely a better candidate. Figure~\ref{fig:dockerint} displays an example container workflow we find works well for sharing analyses. Here, we conduct analysis with Jupyter and then use Jupytext to mirror the analysis into R Markdown, a code script, and a HTML rendering for showcasing. These files are then containerized by building off a custom base image we have created containing Python, R, Jupyter, R Studio Server, and R Shiny. Once running, the container is accessible through the host computer's web browser where a start-page offers several options to interact with the analysis including browsing the files (e.g. to view the HTML rendering) or opening the notebooks in a graphical interface like Jupyter or RStudio.

\section{Code Sharing and Beyond} \label{sec:concl}

Containerizing code has potential benefits in a wide variety of contexts. For example, in a peer review process journals might ask authors to provide an interactive containerized version of their analysis. Building such an image is relatively easy to do and would allow reviewers to quickly assess a working version of the code. 
Additionally, containerized analyses might provide a more secure way for reviewers to run third party code. For example, when using Docker, containerized code cannot see, change, affect, or in any way alter other containers or the host system. For security-conscious individuals or institutions this may be attractive. An added benefit is that one need not clutter up their system installing single-use libraries in order to evaluate third party analyses. 

Containerized code notebooks may also be used as a tool for teaching allowing distribution of identical code, data, and a computing environments to all students. Conversely, student projects in applied statistical courses could be containerized before submission. While a small amount of time would need to be devoted to teaching students some simple mechanics of containerization, in our estimation this is not more complicated than other coding tasks required in many courses and would provide an opportunity for a discussion with students about research reproducibility, replicability as well as good coding practices.

Beyond the direct benefits of making code more easily shareable, the act of containerizing analyses can itself serve as a helpful review step in a scientific pipeline. Preparing analyses for containerization forces one to review the code. This encourages simplification and refactoring of code, as well as writing of the associated documentation and commentary. If the final results included in a manuscript are the output of a container then one can be ensured the results are computationally reproducible. 
Additionally, software like Docker can be interwoven seamlessly into popular code sharing and versioning workflows. For example, one can connect github and Dockerhub accounts together so that updates to code on github are automatically propagated to Dockerhub where an image is subsequently built. Alternatively, Docker can directly pull and build repositories from github. 

Containerization is more than just an approach for preserving passive code archives. It allows rich interaction and exploration of analysis and helps create  usable and reproducible analyses. Containerizing interactive analyses can enhance the ability of statisticians to easily share code, analyses, and ultimately ideas.

\bibliographystyle{chicago}

\bibliography{Reproducibility}
\end{document}